\documentstyle{aipproc}

\begin{document}
\title{Berry phase induced persistent current in mesoscopic systems}

\author{Shiro Kawabata\\
Physical Science Division, Electrotechnical Laboratory\\
1-1-4 Umezono, Tsukuba,  Ibaraki 305-8568, Japan}


\maketitle

\bigskip
\bigskip
\bigskip
\bigskip

Since the discovery of the Berry phase~[1], there has been much interest in the study of topological effects in 
the fields of quantum mechanics and condensed matter physics~[2].
The typical example to illustrate the Berry phase is the Aharonov-Bohm (AB) effect~[3] in the mesoscopic ring~[4$\sim$10],
where the relative phase would accumulate on the wave function of a charged particle due to the presence of a electromagnetic gauge potential.
Similarly, when a quantum spin follows adiabatically a magnetic field that rotates slowly in time, the spin wave function 
acquires an additional geometric phase (Berry phase) besides the usual electromagnetic phase in the static magnetic fields.

In this paper we investigate the persistent current~[11$\sim$17] of the  quasi-one-dimensional disordered rings in the presence of a static 
inhomogeneous magnetic field and show that the spin wave function accumulates the Berry phase when 
the spin of an electron traversing an AB ring adiabatically follows an inhomogeneous magnetic filed with a tilt angle 
and this phase leads to persistent equilibrium current~[18].

We begin by considering a quasi-one-dimensional ring of circumference $L_x=2 \pi r$ and volume $V=L_xL_yL_z$.
The ring is embedded in an static inhomogeneous  magnetic field ${\bf B}$.
For a spin-$1/2$ electrons of mass $m$ and charge $e$, the system may be described by the Hamiltonian
\begin{eqnarray}
{\cal H}
      =
      \frac{1}{2m} \left[  {\bf p} - \frac{e}{c} {\bf A}^{em}({\bf r}) \right]^2 
     +u({\bf r} ) 
	 -\frac{1}{2} g \mu_B {\bf B} ({\bf r} ) 
	 \mbox{\boldmath $\cdot$}
	 \mbox{\boldmath $\sigma$}
	 ,
  \label{eqn:e1}
\end{eqnarray}
where ${\bf p}$, ${\bf r}$, $g$, $\mu_B$, and $\hbar \mbox{\boldmath $\sigma$}/2$ are the momentum, position, $g$ factor, Bohr magneton, and spin, respectively.
The operator $u({\bf r})$ represents the spin-independent random impurity potential and, ${\bf A}^{em}$ is the electromagnetic gauge potential, with ${\bf B}=\nabla \times {\bf A}^{em}$ relating it to the magnetic fields.
In the following we specialize in the case of inhomogeneous magnetic fields with constant magnitude $B$, and we have parametrized ${\bf B}$ in terms of the spherical polar angles $\chi$ and $\eta$ so that it has Cartesian components $B\left(  \sin \chi({\bf r}) \cos \eta({\bf r}),  \sin \chi({\bf r}) \sin \eta({\bf r}), \cos \chi({\bf r}) \right)$, with the angles $\chi$ and $\eta$ being smooth functions of position.

Using the Green's function, the canonical disorder-averaged persistent current is given by
\begin{equation}
      \left< I(\Phi^{em}) \right>
	  \approx 
	  - \frac{\Delta V^2}{2} 
         \frac{ \partial}{\partial \Phi^{em}} 
	  \int_{-\infty}^{\infty} d \varepsilon_1 
	  \int_{-\infty}^{\infty} d \varepsilon_2
	  f(\varepsilon_1 )  f(\varepsilon_2 )
	  \sum_{\alpha,\alpha'} 
	  K_{\alpha,\alpha'} (\varepsilon_1,\varepsilon_2)
		 ,
  \label{eqn:e3}
\end{equation}
where  $\Delta $,   $\Phi^{em}$, $\alpha$ and $f$ are the mean level spacing, the spin index, the electromagnetic flux and the 
Fermi-Dirac distribution function., respectively.
In this equation,  the two-point correlator of the density of state $K_{\alpha,\alpha'}$ is defined as
\begin{eqnarray}
	  K_{\alpha,\alpha'} (\varepsilon_1,\varepsilon_2)
	  =
	  \frac{1}{2 \pi^2 V^2 \hbar^2}
	  \mbox{Re}
	  \int d {\bf x}_1
	  \int d {\bf x}_2
	  \;
	  {\cal C}_{\alpha,\alpha'} 
	  \left( {\bf x}_1,{\bf x}_2;\varepsilon_1-\varepsilon_2\right)
	  {\cal C}_{\alpha,\alpha'} 
	  \left( {\bf x}_2,{\bf x}_1;\varepsilon_1-\varepsilon_2\right)
		.
  \label{eqn:e7}
\end{eqnarray}
Here we have used the definition of the particle-particle pair propagator 
\begin{eqnarray}
	  {\cal C}_{\alpha,\alpha'} 
	  \left( {\bf x}_1,{\bf x}_2; \varepsilon_1-\varepsilon_2\right)
	  =
      \frac{2 \pi \rho(0)}{\hbar}
	  \left<
	           G_{\alpha,\alpha}^{R} ({\bf x}_2,{\bf x}_1;\varepsilon_1)
	           G_{\alpha',\alpha'}^{A} ({\bf x}_2,{\bf x}_1;\varepsilon_2)
	  \right>
		,
  \label{eqn:e8}
\end{eqnarray}
where $G^{R(A)}$ is the retarded (advanced) Green's function and $\rho(0)$ is the density of states (per unit volume and spin) at the Fermi surface.
We have evaluated the pair propagator $ {\cal C}_{\alpha,\alpha'}$  by using the diagrammatic method~[19,20] and obtained
\begin{eqnarray}
 	 \left< I (\Phi^{em}) \right>
	 =
	 - \frac{\Delta}{2 \pi \beta}
	 \sum_{\alpha=\pm 1} \mbox{Re} \frac{\partial}{\partial \Phi^{em}}
	 \sum_{\nu_\ell>0}  
	 \sum_{n_x=-\infty} ^{\infty}
	 \frac{\nu_\ell}{
	                \left\{
					\displaystyle{
	                \nu_\ell
					+
					\frac{\hbar}{\tau_\varphi}
					+
					4 \pi^2 E_{Th}
					\left( n_x + 2 \Phi_\alpha \right)^2
					}
					\right\}^2
					}	
					,
  \label{eqn:e17}
\end{eqnarray}
where $\beta=1/k_BT$ and $E_{Th}=\hbar D/L_x^2$ is the Thouless energy, $\nu_\ell=2\pi \ell /\beta$ ($\ell$ is an integer) is the 
boson Matsubara frequency and $\tau_\varphi=L_{\varphi}^2/D$ ($L_{\varphi}$ is the phase coherence length) is the phase coherence time, respectively.
In this equation $\Phi_{\alpha} = \Phi^{em}/\Phi_0 + \alpha \Phi^g$, where $\Phi^{em}$ is the electromagnetic flux through the area $\pi r^2$ and the 
geometric flux $\Phi^g$ which corresponds to the Berry phase is given by
\begin{eqnarray}
   \Phi^g
   =
   \frac{1}{4 \pi} 
   \int_0^{2 \pi}
   d \phi
   \left[ 
    \cos \chi (\phi)
	-
	1
   \right]
   \partial_{\phi} \eta(\phi)
   .
  \label{eqn:e13}
\end{eqnarray}
The Berry phase arises from adiabatic approximation for the spin (dynamical Zeeman) 
propagator.
It should be note that the expression Eq.~(\ref{eqn:e17}) is only valid in the adiabatic regime in which the spin 
of the electron adiabatically follows the local direction of the non-uniform magnetic field.
This adiabaticity requires that the precession frequency $\omega_B=g\mu_B B / 2 \hbar$ is large compared to 
the reciprocal of the diffusion time $\tau_d=L_x^2/D$ 
($D$ is the diffusion constant) around the ring, i.e., $\omega_B \tau_d \gg 1$, or equivalently $B \gg B_c \equiv 2E_{Th}/g \mu_B$~[21$\sim$23].

In the limit of zero temperature, the Matsubara sum turns into an integral $\sum\nolimits_{\nu} \to 2 \pi/\beta  \int d \nu$, which is 
easily evaluated.
This yields for the averaged persistent current at $T=0$ as
\begin{eqnarray}
 	 \left< I (\Phi^{em}) \right>
	 &=&
	 \frac{I_0}{M}
	 \sum_{\alpha=\pm 1} 
	 \sum_{n=1}^{\infty} 
	 \exp \left( 
	                      -n \frac{L_x}{L_{\varphi}} 
	           \right)
	 \sin \left( 4 \pi n \Phi_\alpha \right)
	 \nonumber\\
	 &=&
		 \frac{I_0}{2M}
	 \sum_{\alpha=\pm 1} 
	 \frac{
	             \displaystyle{
	             \sin \left(  4 \pi \Phi_\alpha \right)
				 }
	           }
			   {
			    \displaystyle{
				\cosh \left( \frac{L_x}{L^B_{\varphi}} \right)
				-
				\cos \left(  4 \pi \Phi_\alpha \right)
				}
				}
	        .
  \label{eqn:e18}
\end{eqnarray}
$I_0=e \upsilon_F/L_x$ is the current carried by a single electron state in an ideal one-dimensional ring and  $M=k_F^2V/L_x$ is the 
effective channel number, where $\upsilon_F$ is the Fermi velocity and $k_F=m\upsilon_F/\hbar$ is the Fermi wave number.
The average current  is a periodic function of $\Phi^{em}$ and $\Phi^g$ with period $\Phi_0/2$ and $1/2$, respectively.
As the half flux periodicity of the electromagnetic flux in disordered rings~[14], that of geometric flux is ascribed to 
the ensemble averaging for fixed particle number: averaging eliminates the first Fourier component of the current although the 
second components which results from the interference between time-reversed trajectories survives.

In summary we have investigated the effect of Berry phase on the persistent current in a static inhomogeneous magnetic field and showed that
the disorder-averaged current oscillates as a function of the geometric flux.


%
%
%
%
%
%
%
%
%
%

%
 
\end{document}